# Earthquake Disaster based Efficient Resource Utilization Technique in IaaS Cloud

Sukhpal Singh, Rishideep Singh

*Abstract*— Cloud Computing is an emerging area. The main aim of the initial search-and-rescue period after strong earthquakes is to reduce the whole number of mortalities. One main trouble rising in this period is to and the greatest assignment of available resources to functioning zones. For this issue a dynamic optimization model is presented. The model uses thorough descriptions of the operational zones and of the available resources to determine the resource performance and efficiency for different workloads related to the response. A suitable solution method for the model is offered as well. In this paper, Earthquake Disaster Based Resource Scheduling (EDBRS) Framework has been proposed. The allocation of resources to cloud workloads based on urgency (emergency during Earthquake Disaster). Based on this criterion, the resource scheduling algorithm has been proposed. The performance of the proposed algorithm has been assessed with the existing common scheduling algorithms through the CloudSim. The experimental results show that the proposed algorithm outperforms the existing algorithms by reducing execution cost and time of cloud consumer workloads submitted to the cloud.

## I. INTRODUCTION

Cloud Computing discusses both the application provided as services over the Internet and the hardware and systems software in the datacenters that offer those facilities. The facilities themselves have long been referred to as Software as a Service (SaaS) [1]. The datacenter hardware and software is what we will call a Cloud. Cloud is made accessible in a pay-as-you-go manner to the cloud consumers. Private Cloud to denote to internal datacenters of a business or other organization not made accessible to the general public. Thus, Cloud Computing is the combination of SaaS and Utility Computing, but does not contain Private Clouds [2]. Infrastructure as a Service (IaaS) is the supply of hardware (server, storage and network), and associated software (OS virtualization technology, file system), as a service. It is a progress of traditional hosting that does not want any long term guarantee and permits cloud consumers to provision resources on demand. The IaaS provider does very little supervision other than retain the data centre operational and cloud consumers must install and maintain the software services them just the technique they would in their own data center [3].

To make effective use of the remarkable abilities of the cloud, efficient scheduling algorithms are mandatory. These scheduling algorithms are generally applied by cloud resource manager to optimally dispatch workloads to the cloud resources. There are comparatively a large number of scheduling algorithms to reduce the total execution time of the workloads in distributed systems [4]. Really, these algorithms attempt to reduce the overall execution time of the workloads by finding the most appropriate resources to be allocated to the workloads. It should be observed that decreasing the overall execution time of the workloads does not essentially result in the reduction of execution time of each distinct workload [5].

Earthquakes happen when the earth's tectonic plates release stress. This release of pressure generates the sensations, which we call an earthquake, causing damage to the surrounding atmosphere. The strength of earthquakes is measured on the Richter scale [6]. To avoid the earthquake situation, provide resources quickly to the affected people based on deadline urgency. The same concept has been used in our proposed algorithm.

The rest of the paper is organized as follows. Related work has been presented in Section 2. In Section 3, a description of the Earthquake Disaster Based Resource Scheduling (EDBRS) Framework has been presented. Experimental Results and Discussion has been presented in Section 4. Conclusions and the future works have been presented in Section 5.

## II. RELATED WORK

Old-fashioned way for scheduling in cloud computing tended to use the direct tasks of cloud consumers as the overhead workload base. The problem is that there may be no relationship among the overhead workload base and the way that different tasks cause overhead costs of resources in cloud systems [7]. For large number of simple cloud workloads this upturns the cost and the cost is reduced if we have small number of complex cloud workloads.

Ke Liu et al. [8] presented an innovative compromised-time-cost (CTC) scheduling algorithm which considers the features of cloud computing to accommodate instance-intensive cost-constrained workflows by compromising execution time and cost with cloud consumer input allowed on the fly. The simulation based on SwinDeW-C has verified that CTC algorithm can attain

*Manuscript received May, 2013*.
 *Sukhpal Singh*, *Computer Science and Engineering Department, Thapar University, (e-mail: ssgill@hotmail.co.in), Patiala, India, Mobile No. +91-81462-08007*
 *Rishideep Singh*, *HOD, Department of Information Technology, N.W.I.E.T. Dhudike, Moga, India, Mobile No. +91-98145-73076, (e-mail: rishideeps@yahoo.co.in).*

lesser execution cost than others while meeting the consumer-designated deadline or decrease the mean execution time than others within the consumer-designated execution cost [8].

Suraj Pandey et al. [9] presented a particle swarm optimization (PSO) based heuristic to schedule cloud workloads to resources that takes into account both communication cost and data transmission cost. It is used for workflow cloud workload by changing its computation costs. PSO can attain cost savings and good distribution of cloud workload onto resources [20].

Selvarani et al. [10] proposed an improved cost-based scheduling algorithm for making well-organized mapping of cloud workloads to available resources in cloud. The inventiveness of traditional activity based costing is proposed by new task scheduling strategy for cloud environment where there may be no relation among the overhead workload base and the way that different tasks cause overhead cost of resources in cloud. This scheduling algorithm splits all consumer workloads depending on priority of each task into three dissimilar queues. This scheduling algorithm calculates both resource cost and computation performance, it also Increases the computation to communication ratio [21].

Saeed Parsa et al. [11] proposed a new task scheduling algorithm RASA, combination of Max-min and Min-min. RASA uses the benefits of Max-min and Min-min algorithms and shelters their disadvantages. However the deadline of each workload, arriving rate of the workloads, cost of the workload execution on each of the resource, cost of the communication are not considered. The experimental results show that RASA is overtakes the existing scheduling algorithms in large scale distributed systems [22].

Yun Yang et al. [12] proposed a scheduling algorithm which takes cost and time. The simulation has validated that this algorithm can achieve lower cost than others while meeting the user nominated deadline [23]. Cui Lin et al. [13] proposed an SHEFT workflow scheduling algorithm to schedule a workflow elastically on a Cloud computing situation. SHEFT not only outperforms several illustrative workflow scheduling algorithms in optimizing workflow execution time, but also permits resources to scale elastically at runtime [24].

Meng Xu et al. [14] worked on multiple workflows and multiple QoS. They have a policy implemented for multiple workflow management system with multiple QoS. The scheduling access rate is improved by using this policy. This policy decreases the make span and cost of workflows for cloud computing platform [25].

Scheduling is one of the main problems in the management of workload execution in cloud environment. In this literature survey, we have surveyed the various existing scheduling algorithms in cloud computing. We also observed that execution time is critical in virtual environments. Existing scheduling algorithms does not consider execution cost. Consequently there is a need to implement a scheduling algorithm that can improve the execution time and cost in cloud environment [15].

## III. EARTHQUAKE DISASTER BASED RESOURCE SCHEDULING (EDBRS) FRAMEWORK

Cloud scheduling is the core of the Cloud resource management systems. It essentially suggests mapping Cloud Workloads to the available ingredient resources. This process contains probing multi administrative domains to use the available resources from the Cloud infrastructure in order to fulfill the desires of the user. Cloud scheduling is a two-step process. In step one, the required set of resources is recognized as per the cloud consumer's requests and in the second step, the Cloud Workloads are mapped on to the actual set of resources, thus further ensuring near optimal contentment of QoS parameters. Fig. 1 shows an Earthquake Disaster Based Resource Scheduling (EDBRS) Framework.

### A. Problem Statement

To discover the best resource to a corresponding workload is a tedious task and the problem of finding the best resource–workload pair according to cloud consumer's workload requirements is a combinatorial optimization problem. The main goal of the Cloud scheduler is to schedule the resources effectively and efficiently. The resources and workloads/Cloud Workloads can leave and join the Cloud dynamically.

### B. Resource Scheduling Rules

As The Resource Scheduling problem model of a complex resource allocation system can usually be described as follows: n workloads are executed on m resources, each workload involves q operations and every operation should be executed on an appropriate resource. The following are the suppositions for the scheduling model:

- Every resource can only process one workload at a time.
- The operations cannot be interrupted once beginning.
- There are no priority restrictions among operations of different workloads.
- There are priority restrictions among operations of the same workload.
- The processing time of one operation on a resource is determined.
- At the beginning, all workloads can be processed.
- Communication time is not considered.

The scheduling objective is to meet the desires of delivery date, decrease the total processing time and maximize the utilization of resources as much as possible. Compared to other resource scheduling algorithms, this algorithm is more appropriate for actual implementation. For parallel resource problems, we not only consider the circumstances of multiple resources being capable of executing one operation, but also consider the situation of different operations competing for one resource.

### C. Optimal Resource Assignment

In this section, the multi-cloud model and two dissimilar kinds of BoW (Equal and variable length workloads) and objective function is described.

- Equal Length Workload: All the workloads have identical processing time.

- Variable Length Workload: All the workloads have different processing time.

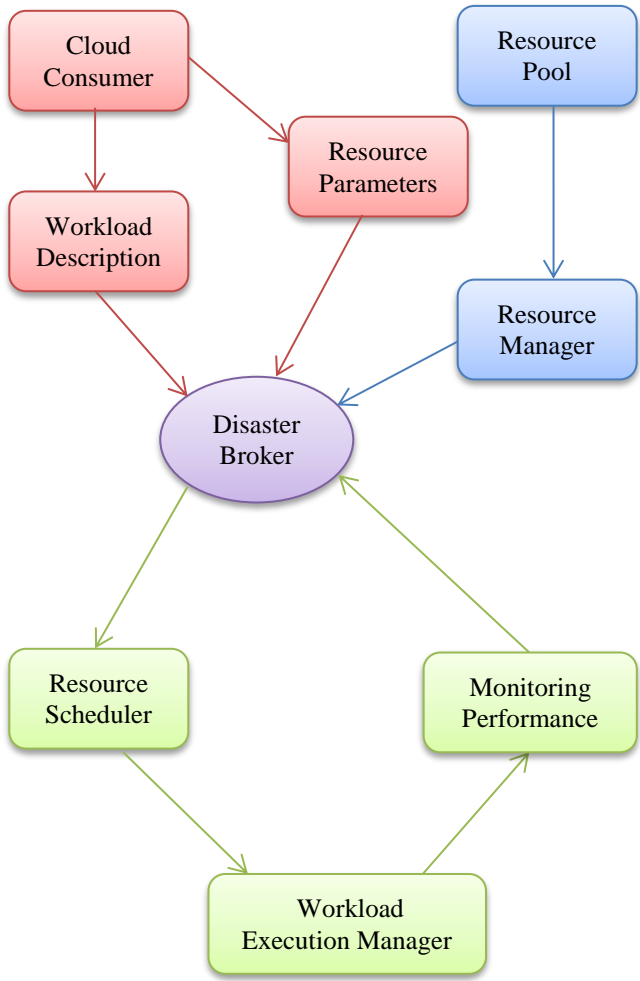

Fig. 1 EDBRS Framework

*1) Multi Cloud Model*: There are exactly k different resource types in the public cloud and their corresponding speed and cost are given by $< SP_1, SP_2 \ldots\ldots\ldots SP_h >$ and $< CO_1, CO_2 \ldots\ldots\ldots CO_h >$ respectively [16]. The price of leasing a resource (irrespective of resource type) is usually charged as pay per use, more accurately as ATU (Accountable Time Unit). The private cloud in this work comprises of homogeneous resources (one type) with the speed of $< SP_f >$ and cost of $< CO_f >$. Suppose that for every resource kind in the public cloud, the number of resources to be rented by a particular cloud customer is limited to $< LE_1, LE_2 \ldots\ldots\ldots LE_h >$ and this relates to the private cloud (i.e., $LE_f$), although $LE_f$ is most probable the total number of private resources (f) [16].

*2) BoW Workload Model:* A BoW comprises of a set of d workloads and each of them is related with an execution time $E_r$ that is known at the time of workload assignment. Since we assume that these workloads are independent with each other and CPU-intensive, all of them are prepared to start at time t = 0 irrespective of kind of cloud. In other words, a workload can be allocated to run either in public cloud resource or in private cloud resource with slight or no deferral in transmission. If workload r is allocated to deploy in a resource of speed $SP_m$, and cost of $CO_m$, then the processing time and the cost of executing workload r is equal to $\frac{E_m}{SP_m}$ and $\frac{CO_m \times E_r}{SP_m}$ respectively. As a cloud customer rents a public cloud resource in whole hour duration, the real price to complete a workload in a particular public cloud resource is equal to

$$CO_m \left\lceil \frac{\sum_{r \epsilon R_m}^d E_r}{SP_m} \right\rceil$$

Where $R_m$ is a set of all workloads assigned to resource m.

*3) Objective Function*: An objective function that goals to decrease the sum of product of cost and time expended for finishing all workloads of a given BoW. This objective function successfully captures the compromise between performance improvement and additional price related with public resource payment. Further formally, the workload assignment problem with the cost function of each resource can be generally formulated as an Integer Program as follows:

$$min\ z = \sum_{m \epsilon \Gamma \cup \{f\}} Execution\ Cost_m \times Execution\ Time_m$$

where $\Gamma$ is a set of all resources offered in the public cloud for scheduling. $\Gamma$ might be a certain number of public resources the cloud consumer chooses to lease or it is restricted to resource limit set by the public cloud provider.

*4) Scaling out BoW across Multiple Clouds*: Offer a cloud customer-side scheduler that discovers cost effective solution (optimal or at least near-optimal) for running BoW across both public and private cloud resources. Optimal result procedures for models that compromise between time and cost (may be with desired deadline and/or estimated budget constraints) are extremely striking in exercise. Awkwardly, these models are computationally inflexible, and creating near-optimal polynomial-time heuristics for them is extremely challenging. The origination of workload assignment problems with equal-length and varying-length workloads have been offered and offer a unique solution algorithm for each of these two problems.

*5) Equal-length workloads*: Offer properties of the optimal assignment of d workloads $r_1, r_2, \ldots\ldots\ldots r_d$ with the similar execution time E on a set ($\Gamma$) of public cloud resources of h different types and a set (${f}$) of private cloud resources of the similar kind. More formally, the problem can be indicated as an integer programming one as follows:

$$min\ z = \sum_{m \epsilon \Gamma}^{\infty} \left( LE_m CO_m \left\lceil \frac{q_m E}{SP_m} \right\rceil \frac{q_m E}{SP_m} \right.$$
$$\left. + LE_f\ CO_f\ \left(\frac{q_f E}{SP_f}\right) \right)$$

$$\sum_{m \epsilon \Gamma} LE_m q_m + LE_f CO_f = d$$

$$q_f, q_m\ \epsilon\ \mathbb{Z} \geq 0$$

1935

Where $q_m$ is the number of workloads allocated to run on resources of kind m. $q_m$ is calculated using above Equations.

*6) Accurately a Varying length workloads*: In this, the properties of an optimum assignment of d workloads with unlike processing times. Suppose that the execution time of each workload relates to only one of the predetermined values A = {$E_1, E_2, \ldots\ldots E_A$}. Let $d_a$ shows the number of workloads per every group a; $\forall$ 1≤ a ≤|A|. Again, there are a set of Γ of public cloud resources of h dissimilar forms and a set of private cloud resources of the similar kind. The objective function is to reduce the sum of product of cost and time of finishing all workloads on all available resources can be stated officially as follows:

$$\min z =$$

$$\sum_{m\epsilon\Gamma} LE_m CO_m \lceil \frac{\sum_{1\leq a\leq A} q_{m,a}E_a}{SP_m} \rceil \frac{\sum_{1\leq a\leq A} q_{m,a}E_a}{SP_m}$$

$$+LE_f CO_f(\frac{\sum_{1\leq a\leq A} q_{m,a}E_a}{SP_f})$$

With following conditions

$$\sum_{m\epsilon\Gamma} LE_m \sum_{1\leq a\leq A} q_{m,a} + LE_f \sum_{1\leq a\leq A} x_{f,g} = d_g$$

$$q_{f,a}, q_{f,a} \ \epsilon \ \mathbb{Z} \geq 0 \ , \forall m \ \epsilon \ \Gamma, 1 \leq a \leq |A|$$

In the above formulas, $q_{m,a}$ and $q_{m,a}$ are the number of workloads with particular size of $E_a$ allocated to run on resources of kind m, correspondingly.

*D. EDBRS Based Resource Scheduling Algorithm*

First of all, sort all Cloud Workloads which need to be scheduled according to some rules, and then divide the sorted Cloud Workloads into some batches. Finally, schedule all batches, respectively. Using batch scheduling can guarantee that the Cloud Workload with earlier delivery date has higher priority to occupy resources and can be scheduled earlier. In this algorithm, the strategy of minimizing waiting time is adopted, which means after processing one operation, the next operation will start as soon as possible. In this way waiting time among different operations is shortened, meanwhile the operations of the Cloud Workload with higher priority can be scheduled earlier. In the sorted operation list, the Cloud Workload with higher priority should be scheduled first. If one operation is the first operation of a Cloud Workload, a resource which will be idle the earliest should be selected. If the operation is not the first one, the previous operation's completion time should be checked and there are three cases: firstly, at the completion time of the previous operation there is a resource which can be used to process this operation; secondly, a resource has to wait until the previous operation is completed; and thirdly, after completing the previous operation, the Cloud Workload has to wait until some resource becomes idle. Fig. 2 shows EDBRS Based Resource Scheduling Algorithm.

| EDBRS Based Resource Scheduling Algorithm |
|---|
| 1. Sort all Cloud Workloads<br>The sorting guidelines are as follows:<br>  i. For different Cloud Workloads<br>    a) Sort all the Cloud Workloads according to delivery date, with earlier delivery dates ahead of later ones.<br>    b) The Cloud Workloads with the similar delivery dates are should be sorted by the required amount of Cloud Workloads.<br>    c) The Cloud Workloads with the similar delivery date and the same required amounts should be sorted by the total processing time of one Cloud Workloads.<br>    d) The Cloud Workloads with the same delivery dates, the same required amounts and the same processing time should be sorted by the number of operations of one Cloud Workloads.<br>  ii. For one Cloud Workloads<br>    The different operations of one Cloud Workloads should be sorted by constraints.<br>  iii. According to the processing cycle, all Cloud Workloads should be distributed into some batches.<br>2. Scheduling sorted operations<br><br>**STEP1:** If an operation is the first one of Cloud Workload, the resource which will be idle the earliest in the Cloud Workload's resource list should be selected and the operation will be processed on this resource.<br>**STEP2:**<br>  A. If the operation is not the first one of the Cloud Workload, the previous operation's completion time should be checked. If at that time, there is a resource can be used then schedule this operation to occupy this resource.<br>  B. If there is no operation and resource satisfying case (A), a resource which will be idle earlier than the previous operation's completion time, and whose waiting time is the shortest should be selected.<br>  C. If there is no operation and resource satisfying case (A) and case (B), a resource which will be idle later than the previous operation's completion time should be selected and the operation should be scheduled to occupy this resource. |

Fig. 2 A EDBRS Based Resource Scheduling Algorithm

IV. EXPERIMENT RESULTS AND DISCUSSIONS

In this section, we have defined the performance evaluation criteria to evaluate the performance of an EDBRS Based Resource Scheduling Algorithm. We have selected two matrices, namely execution time and cost for evaluating the performance. The former indicates the total execution time whereas the latter indicates the cost per unit resource that is consumed by the cloud consumers for the execution of their workloads. The execution time and cost are measured in seconds and dollars ($) respectively. To validate our algorithm, 300 cloud workloads and 50–70 resources have been considered. We have used an average of fifty runs in order to guarantee statistical correctness. We have presented the simulation result using Cloudsim [17] so as to test the

performance of the EDBRS Based Resource Scheduling Algorithm. In addition, a comparison of execution time and cost of the proposed algorithm with existing algorithms i.e. OATSB-ABC [18] has been presented. To evaluate the performance of the proposed approach, we have investigated the effects of different numbers of workloads. We have also performed experiments to determine the effect of an increase in the number of workloads on cost and execution time. From the experimental results shown in Fig. 3, we can conclude that the time taken to execute a workload reduces by using the EDBRS Based Resource Scheduling Algorithm.

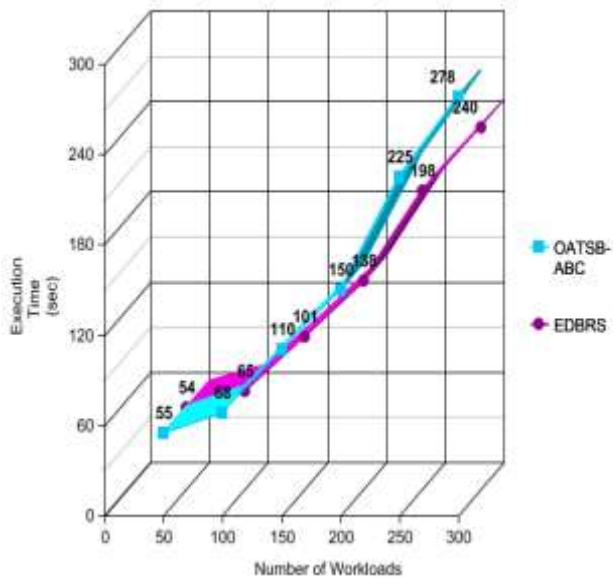

Fig. 3 Comparison of Execution time

Fig. 4 shows that cost per workload increases as the number of submitted workloads increases. The existing algorithm based workload's execution resulted in a schedule which is expensive in comparison to the EDBRS Based Resource Scheduling Algorithm. From all the experimental results, we observed that workload execution using the EDBRS Based Resource Scheduling Algorithm provides the following advantages. The execution time is much lower in comparison to the OATSB-ABC.

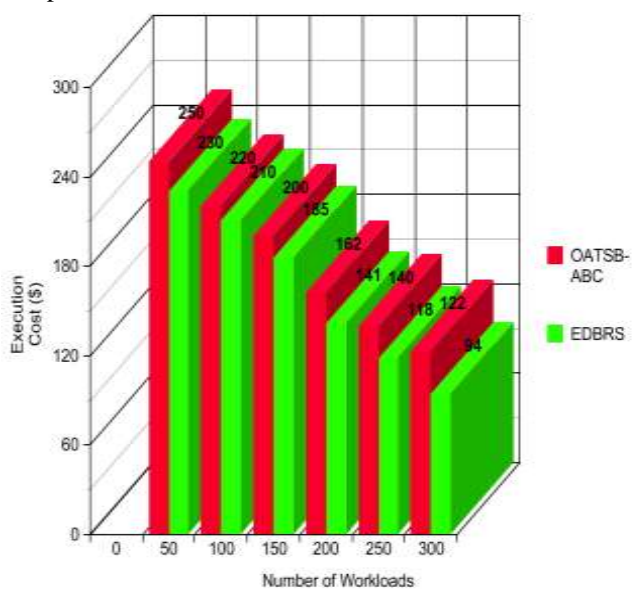

Fig. 4 Comparison of Execution time

We have compared the performance of the EDBRS Based Resource Scheduling Algorithm with well-known scheduling algorithm such as OATSB-ABC. We have analysed the performance of the proposed algorithm with variation in both the number of workloads and the number of resources, which are expected to vary in the real Cloud environment. We evaluated the algorithm's performance with respect to execution time and cost. Execution time allows the evaluation of the algorithm which results in better scheduling in the sense of the duration of workload execution, while the cost allows the comparison for resource selection. The proposed algorithm helps to achieve high performance and simultaneously it also helps to satisfy the cloud consumer's requirements. In the experiments conducted, the EDBRS Based Resource Scheduling Algorithm clearly demonstrates its ability to provide better performance with respect to the existing Cloud scheduling algorithms.

## V. CONCLUSIONS AND FUTURE SCOPE

As Clouds have become more prevalent for processing of large amounts of data, techniques for efficiently utilizing their resources become increasingly significant. The problem of resource provisioning and scheduling is crucial not only to achieve high Cloud performance, but also to satisfy various cloud consumer's demands in an equitable fashion. In this paper, we have proposed an EDBRS Based Resource Scheduling Algorithm for scheduling of independent parallel workloads in the Cloud environment so as to simultaneously minimize the cost and the execution time. We have compared the proposed algorithm with existing heuristic based algorithm. The experimental results show that the EDBRS Based Resource Scheduling Algorithm outperforms in all the cases. The proposed algorithm not only minimizes cost but it also minimizes the execution time. A Cloud is dynamic and diverse, which is rendered by various workloads and resources. This paper mainly focuses on the resource scheduling problems in the Cloud computing environment. In future, we would like to incorporate trust of node and reliability of the node /resources at the time of scheduling of the resources. Current results have been gathered through simulation on CloudSim but in future the same results would be verified on actual Cloud resources.

REFERENCES

[1] Armbrust, Michael, Armando Fox, Rean Griffith, Anthony D. Joseph, Randy Katz, Andy Konwinski, Gunho Lee et al. "A view of cloud computing." Communications of the ACM 53, no. 4 (2010): 50-58.
[2] Foster, Ian, Yong Zhao, Ioan Raicu, and Shiyong Lu. "Cloud computing and Cloud computing 360-degree compared." In Cloud Computing Environments Workshop, 2008. GCE'08, pp. 1-10. Ieee, 2008.
[3] Singh, Sukhpal, and Inderveer Chana. "Introducing Agility in Cloud Based Software Development through ASD." International Journal of u- and e- Service, Science and Technology, Vol.6, No.5, 2013, pp.191-202, ISSN: 2005-4246.
[4] Bhardwaj, Sushil, Leena Jain, and Sandeep Jain. "Cloud computing: A study of infrastructure as a service (IAAS)." International Journal of engineering and information Technology 2, no. 1 (2010): 60-63.
[5] Li, Qiang, and Yike Guo. "Optimization of resource scheduling in cloud computing." In Symbolic and Numeric Algorithms for Scientific Computing (SYNASC), 2010 12th International Symposium on, pp. 315-320. IEEE, 2010.
[6] Zhong, Hai, Kun Tao, and Xuejie Zhang. "An Approach to Optimized Resource Scheduling Algorithm for Open-source Cloud Systems." In

[7] Khattri, K. N. "Great earthquakes, seismicity gaps and potential for earthquake disaster along the Himalaya plate boundary." Tectonophysics 138, no. 1 (1987): 79-92.
[8] Singh, Sukhpal, and Rishideep Singh. "Reusability Framework for Cloud Computing." International Journal of Computational Engineering Research (ijceronline.com) Vol. 2 Issue. 6, ISSN 2250-3005(online), October| 2012, 169-177.
[9] Rimal, Bhaskar Prasad, Eunmi Choi, and Ian Lumb. "A taxonomy and survey of cloud computing systems." In INC, IMS and IDC, 2009. NCM'09. Fifth International Joint Conference on, pp. 44-51. Ieee, 2009.
[10] Liu, Ke, Hai Jin, Jinjun Chen, Xiao Liu, Dong Yuan, and Yun Yang. "A compromised-time-cost scheduling algorithm in swindew-c for instance-intensive cost-constrained workflows on a cloud computing platform." International Journal of High Performance Computing Workloads 24, no. 4 (2010): 445-456.
[11] Pandey, Suraj, Linlin Wu, Siddeswara Mayura Guru, and Rajkumar Buyya. "A particle swarm optimization-based heuristic for scheduling workflow workloads in cloud computing environments." In Advanced Information Networking and Workloads (AINA), 2010 24th IEEE International Conference on, pp. 400-407. IEEE, 2010.
[12] Selvarani, S., and G. Sudha Sadhasivam. "Improved cost-based algorithm for task scheduling in cloud computing." In Computational Intelligence and Computing Research (ICCIC), 2010 IEEE International Conference on, pp. 1-5. IEEE, 2010.
[13] Parsa, Saeed, and Reza Entezari-Maleki. "RASA: A new task scheduling algorithm in Cloud environment." World Applied sciences journal 7 (2009): 152-160.
[14] Yuan, Dong, Yun Yang, Xiao Liu, and Jinjun Chen. "A data placement strategy in scientific cloud workflows." Future Generation Computer Systems 26, no. 8 (2010): 1200-1214.
[15] Singh S, Chana I (2013) Cloud based development issues: a methodological analysis. Int J Cloud Comput Serv Sci 2(1):73–84
[16] Lin, Cui, and Shiyong Lu. "Scheduling scientific workflows elastically for cloud computing." In Cloud Computing (CLOUD), 2011 IEEE International Conference on, pp. 746-747. IEEE, 2011.
[17] Xu, Meng, Lizhen Cui, Haiyang Wang, and Yanbing Bi. "A multiple QoS constrained scheduling strategy of multiple workflows for cloud computing." In Parallel and Distributed Processing with Workloads, 2009 IEEE International Symposium on, pp. 629-634. IEEE, 2009.
[18] Lodha, Priya R., and Mr Avinash P. Wadhe. "Study of Different Types of Workflow Scheduling Algorithm in Cloud Computing." International Journal of Advanced Research in Computer Science and Electronics Engineering (IJARCSEE) 2, no. 4 (2013): pp-421.
[19] Sukhpal Singh, Inderveer Chana, "Advance Billing and Metering Architecture for Infrastructure as a Service", International Journal of Cloud Computing and Services Science International Journal of Cloud Computing and Services Science (IJ-CLOSER) 2, no. 2 (2013): 123-133.
[20] Gutin, Gregory, and Abraham P. Punnen, eds. The traveling salesman problem and its variations. Vol. 12. Springer, 2002.
[21] Buyya, Rajkumar, Rajiv Ranjan, and Rodrigo N. Calheiros. "Modeling and simulation of scalable Cloud computing environments and the CloudSim toolkit: Challenges and opportunities." In High Performance Computing & Simulation, 2009. HPCS'09. International Conference on, pp. 1-11. IEEE, 2009.
[22] Sukhpal Singh, Inderveer Chana, "Consistency Verification and Quality Assurance (CVQA) Traceability Framework for SaaS", 3rd IEEE International Advance Computing Conference (IACC-2013), February 2013.
[23] Sukhpal Singh, Harinder Singh, "Case Study Based Software Engineering Project Development: State of Art", International Journal of Scientific Research in Computer Science Applications and Management Studies, Volume 2, Issue 3, May 2013.
[24] Q. Cao, Z.-B. Wei, and W.-M. Gong. An optimized algorithm for task scheduling based on activity based costing in cloud computing. In 3rd International Conference on Bioinformatics and Biomedical Engineering ICBBE 2009., pages 1 –3, June. 2009.
[25] Sukhpal Singh, Inderveer Chana, "Enabling Reusability in Agile Software Development", International Journal of Computer Applications, (0975 – 8887) Volume 50 – No.13, July 2012, 33-40.

**Sukhpal Singh** obtained his B.Tech. (Computer Science and Engineering) Degree from G.N.D.E.C. Ludhiana (Punjab) in 2010. He joined the Department of Computer Science & Engineering at North West Institute of Engineering and Technology, Moga (Punjab) in 2010. Presently he is pursuing Master of Engineering (Software Engineering) degree from Thapar University, Patiala. He has a certification of Cloud Computing Fundamentals includes Introduction to Cloud Computing and Aneka Platform (US Patented) by ManjraSoft Pty Ltd, Australia. He is doing his thesis work in the area of Dynamic Scalability of IaaS Resources in Cloud Computing. His research interests include Image Compression, Software Engineering, Resource Scheduling in Cloud Computing, Operating System, Software Testing & Verification and Database Management System. He is an active member of ACM and IEEE.

**Rishideep Singh** is currently a faculty and Head in the department of Information technology, NorthWest Institute of engineering & technology, Dhudike, Moga. He received his Diploma in Computer Servicing & Maintenance in 1997 and B.E. in Computer Science & Engineering in 2000, respectively from SLIET Longowal. He completed his M.Tech. in Information Technology in 2008. He has worked as Lecturer and Head of PG department of computer science and applications at GHG Khalsa College, Gurusar Sadhar, Ludhiana from 2001 to 2010. He joined NorthWest Institute of engineering & technology, Dhudike, Moga in 2010 and is presently associated with the same Institute.. He has published 11 papers in referred journals of repute. His research interests include routing algorithms and cloud computing.